\documentstyle[12pt,epsf]{ioplppt}

\newcommand \real   {I\kern-1.5mm{R}}
\newcommand \nat    {I\kern-1.5mm{N}}
\newcommand \sign   {\mathop{\rm sign}\nolimits}
\newcommand \thetaF {\mathop{\theta}\nolimits}

\begin{document}
\jl{1}
\title{Training a perceptron by a bit sequence: Storage 
capacity}[Training a perceptron by a bit sequence]
\author{M. Schr\"oder\dag, W. Kinzel\dag\ and I. Kanter\ddag}
\address{\dag Institut f\"ur Theoretische Physik, Universit\"at 
W\"urzburg,
Am Hubland, D-97074 W\"urzburg}
\address{\ddag Department of Physics, Bar-Ilan University,
Ramat-Gan 52900, Israel}

\begin{abstract}
A perceptron is trained by a random bit sequence.
In comparison to the corresponding classification problem, the 
storage capacity decreases to $\alpha_c=1.70\pm0.02$ due to
correlations between input and output bits. The numerical results
are supported by a signal to noise analysis of Hebbian weights.
\end{abstract}

\section {Introduction}

Artificial neural networks are successful in predicting time
series (Weigand \etal 1993). Given a sequence of real numbers, a
multilayer network is able to learn from $N$ consecutive
numbers the following one. After learning a part of the
sequence, the network is able to generalize: If $N$ consecutive
numbers are taken from the part of the sequence which the
network has not learned, the network can predict the following
number to some extent.

Using methods and models of statistical mechanics, training from
a set of examples and generalization of neural networks has been
studied intensively (Hertz \etal 1991, Kinzel \etal 1991, Opper 
\etal 1996).
Most work has been concentrated on perceptrons and binary
classification problems. A set of $N$--dimensional input vectors
is classified by a perceptron. A different perceptron is trained
by this set of examples; after the training process the network
is able to generalize: it has some overlap to the weights of the
perceptron which has generated the examples. If the
classification is not performed by a different perceptron but is
assigned randomly, the network can still learn a certain amount
of examples. The maximum number of examples, which can be
classified by a perceptron, is related to the storage capacity 
of the corresponding attractor networks (Gardner 1988).

Only recently this approach has been extended to time series
analysis (Eisenstein \etal 1995)   A perceptron was trained from a
series of bits which was produced by a different perceptron.
Hence also the generation of time series by a nerual network is
interesting in this context, and recently an analytic solution
of a stationary time series generated by a peceptron
has been found (Kanter \etal 1995). 

It turns out that a perceptron can predict bit sequences very
well, if those are taken from stationary time series produced by
a different perceptron (Eisenstein \etal 1995). Already a small
training set leads to perfect prediction of the rest of the
sequence, at least for $N \rightarrow \infty$. However, the
overlap between a learning and a generating network is very
small. 

In this paper we study the analogy of the storage capacity
problem in the context of bit sequences: A set of $P$
consecutive bits, which are randomly chosen, is repeated
periodically (or placed on a ring). A perceptron with $N < P$ is
trained on this bit sequence, where the output bit is given by
the bit which follows the $N$ input bits. Hence, the only
difference to the examples used for the classification problem are
correlations between the input and output: The output bit is
contained in the input of $N$ examples.

In Section 2 we introduce the bit sequence, which we use for
training a perceptron which is simulated in Section 3. Section
4 presents a signal to noise analysis of the Hebbian learning
rule. A general Boolean function is considered in Section 5, and
the last Section contains a summary and the conclusions.

\section {Bit sequence}

$P$ bits $S_i \in \{-1, 1\}; i = 1,\dots, P$ are chosen randomly
and
independently. This sequence is repeated periodically from $i =
- \infty$ to $i=\infty$ (or placed on a ring, equivalently). 
$N$ consecutive
bits are used as an input to a perceptron with weights $w_j \in
\real ; j = 1,\dots, N$ (see Figure 1):
\begin{equation}
\label{eins} \sigma_{\nu} = \sign \sum\limits^{N}_{j=1} \, w_j \,
\xi^{\nu}_j\\
\mbox{with \,\,\,} \xi^{\nu}_j = S_{j - 1 + \nu}
\end{equation}

The problem we are addressing here is the following: Can we find
a weight vector $\underline{w} = (w_1, \dots, w_N)$ which
reproduces the next bits in the sequence, i.\,e.
\begin{equation}
\label{zwei} \sigma_{\nu} = S_{\nu + N} \mbox{\hspace{5mm} for
all
\hspace{5mm}} \nu \in \nat \;.
\end{equation}
In particular we are interested in the maximal number $P_c (N)$
of bits which can be reproduced correctly by a perceptron for
$N \rightarrow \infty$; as usual we define
\begin{equation}
\label{drei} \alpha = P/N \; \; ; \;\; \alpha_c =
\lim_{N \rightarrow \infty} \; \; \frac{P_c (N)}{N} \; .
\end{equation}

There exist mathematical theorems about the number configuration
$\{ \sigma_{\nu}\}$ which can be realized by Eq.(\ref{eins}),
which are already more than 140 years old (Schl\"afli 1950, Cover 1965):
If
the $P$ input vectors $\underline{\xi^{\nu}} = (\xi^{\nu}_1 ,
\dots, \xi^{\nu}_N )$ are in general position; i.\,e. if any
subset of $N$ vectors is linearly independent, then the number
$C(P,N)$ of possible configurations $\{ \sigma_{\nu} \} \in
\{+1, -1\} ^P$ is given by
\begin{equation}
\label{vier} C(P,N) = 2 \sum\limits^{N-1}_{i=0}  \,\, {P-1
\choose i} \,.
\end{equation}

In our case of the random bit sequence we expect the input
vectors $\underline{\xi}^{\nu}$ to be in general position. For
$P \le N$ one obtains $C(P,N) = 2^P$; hence, any bit sequence
with $P \le N$ can be perfectly predicted by a perceptron. For
$P < 2N$ there is still a large fraction of configurations which
is given by Eq.(\ref{eins}); this fraction goes to one for $N
\rightarrow \infty$. This means that for random configurations
$\{ \sigma_{\nu}\}$ the probability to map them by a perceptron
is
one in the limit of $N \rightarrow \infty$. For $P > 2N$ this
probability is zero. Hence, for a perceptron and random examples
one finds $\alpha_c = 2$ (Gardner 1988).

However, in our case the configurations $\{ \sigma_\nu\}$ are
not randomly chosen but taken from the input vectors. Each
output bit $\sigma_{\nu}$ appears in $N$ input vectors
$\underline{\xi}^{\nu + 1},\dots, \underline{\xi}^{\nu + N}$, too.
There are correlations between the input vectors and the output
bits. In addition, only the {\bf fraction} of configuration
$\sigma_{\nu}$ which cannot be reproduced by a perceptron goes
to zero for $N \rightarrow \infty$ and $N < P <2N$; their {\bf
number} is still increasing exponentially with $N$. For
instance, for $N=100$ and $\alpha =1.8$ Eq.(\ref{vier}) gives
about $10^{54}$ configurations which are not linearly separable,
that is 6.7\% of all of the possible $2^{180}$ ones. On the
other side, for $P > 2N$ the number of configurations which can
be reproduced by a perceptron still increases exponentially with
$N$, although their fraction disappears. Hence, it is not
obvious, whether the patterns given by a bit sequence belong to
the first or second class, which means whether $\alpha_c < 2$ or
$\alpha_c >2$.

In the uncorrelated case the storage capacity $\alpha_c$ has
been calculated using the replica method (Gardner 1988). Correlations
between the input vectors do not change the result $\alpha_c =
2$. Only if there is a bias for the output bits {\bf and} for
the input bits the storage capacity $\alpha_c$ increases with
the bias. If the patterns are anticorrelated $\alpha_c$ can be lower
than $\alpha_c =2$, too (L\'opez \etal 1995).

For our problem we have formulated the version space of weights
in terms of replicas. One has to average over $P$ random bits,
only, instead of $P \cdot N$ in the uncorrelated case. However,
we did not succeed in getting rid of the correlations and could
not
solve the integral. Therefore, we have studied the bit sequence
numerically. 

\section {Perceptron: Simulations}
To calculate the storage capacity $\alpha_c$ of the perceptron
being trained by a random bit sequence, we have used two
methods:
\begin{enumerate}
\item 
We have used several routines which try to minimize the number of
errors and indicate whether they did succeed or not. Hence, we
obtained a fraction $f(\alpha, N)$ of patterns for which the
routine could find a solution. The capacity $\alpha_c (N)$ is
defined by $f(\alpha_c, N) = 1/2$. Obviously, we obtain a lower
bound for the true $\alpha_c$, only.
The results did not dependent on the actual algorithm within the 
expected error bounds.

We have used a routine that minimizes the
``linear cost-function'' $E=\sum_{\nu=0}^P\thetaF(1-E^\nu)(1-E^\nu)$
with $E^\nu={1\over N}\sum_{j=1}^N w_j\xi_j^\nu \sigma^\nu$ 
(without constraining the
vector $\underline w$). 

\item
The other estimate uses the median learning time (Priel \etal 1994). For
random patterns the average learning time $\tau_a$ of the
perceptron algorithm diverges as $\tau_a^{-1/2} \sim (\alpha_c
- \alpha)$ for $\alpha \rightarrow \alpha_c$ (Opper 1988). We use
this
power law in our case, too. The median $\tau_m$ of the
distribution of learning times is calculated for $\alpha <
\alpha_c$ and $\alpha_c$ is obtained from a fit to the power law
divergence. This method has the advantage that one does not have
to determine whether a pattern cannot be learned at all. If the
number of learning steps is larger than the median the algorithm
can stop; this saves a large amount of computer time.

\end{enumerate}

Figure 2 and Table 1 show the results of the simulations
\footnote{Preliminary results have been reported 1994 by Bork}.
In the uncorrelated case both of the methods give the exact result
$\alpha_c =2$ within the statistical error and for $N =100$,
already. If we use the input from the bit sequence but random
output bits the results agree with $\alpha_c =2$, too. However,
if in addition we use the output bits from the bit sequence we
obtain $\alpha_c = 1.70 \pm 0.02$. Hence, the correlations
between output bits and input vectors decrease the storage
capacity. For the perceptron it is harder to learn a random bit
sequence than a random classification problem. This is due to
the correlations between input and output but not due to the
correlations between the input vectors.

If a perceptron which has learned a bit sequence perfectly
is used as a bit generator, then any initial state of $N$ bits 
taken from the sequence reproduces the complete sequence.
Hence the sequence is an attractor of the bit generator.
However we found, that the basin of attraction is very small. 
If only one bit is flipped in the initial state then there
is a high probability that the generator runs into a different 
sequence.

We have also studied two additional problems:
\begin{enumerate}
\item
The $P$ random bits are not repeated periodically but the
perceptron is trained with a string of $N+P$ random bits. Hence,
there are still $P$ patterns but an output bit belongs only to
part of the other input patterns. On average the correlations
are weaker. Indeed, we find that the storage capacity $\alpha_c
= 1.82 \pm 0.02$ is larger than the one for the periodic
sequence.

\item 
With a bias $m = \frac{1}{p} \, \sum\limits^{P}_{i=1} \, S_i$ in
the bit sequence, the storage capacity increases. This is
similar to the random classification problem (Gardner 1988).

\end{enumerate}

\section {Perceptron: Hebbian learning rule}

In order to get some insight from analytic calculations we now
consider the Hebbian learning rule
\begin{equation}
\label{fuenf} \underline{w} = \frac{1}{N} \,
\sum\limits^{P}_{\nu =1} \, \sigma_{\nu} \underline{\xi}^{\nu}
\,.
\end{equation}

Output bits $\sigma_{\nu}$ and input vectors
$\underline{\xi}^{\nu}$ are taken from a bit sequence $\{S_i\}$,
Eqs.(\ref{eins}) and (\ref{zwei}). It is known that the Hebbian
weights cannot map the examples perfectly. However, the training
error can be calculated from a signal to noise analysis (see for
instance Hertz \etal 1991). The sign of the following stability $E^{\nu}$
shows whether an example is classified correctly.
\begin{equation}
\label{sechs} E^{\nu} = \sigma^{\nu}
\underline{w}\underline{\xi}^{\nu} =
\frac{1}{N} \, \sum\limits^{N}_{i=1} \, \sum\limits^P_{\mu =1}
\, \sigma^{\nu}\, \sigma^{\mu} \xi^{\nu}_i \xi^{\mu}_i \,.
\end{equation}

The fraction of negative values of $E^{\nu}$ defines the
training error.

We calculated the first two moments $\langle E \rangle$ and
$\langle E^2 \rangle$ of $E^{\nu}$, where $\langle \dots \rangle$
means an average over the distribution of the examples, i.e.
over all realizations of the bit sequence. If all bits
$\sigma_{\nu}$ and $\xi^{\nu}_i$ are random one has
\begin{equation}
\label{sieben} \langle \sigma^{\nu} \sigma^{\mu} \rangle =
\delta_{\nu \mu} \;\; ; \;\; \langle \xi^{\nu}_i \xi^{\mu}_j 
\rangle = \delta_{\nu \mu} \delta_{i\,j} \,.
\end{equation}

This gives 
\[
\langle E \rangle = 1 \;\; ; \;\; \langle E^2 \rangle = 1+
\alpha \,.
\]
In the limit $N \rightarrow \infty$ the values of $E^{\nu}$ are
Gaussian distributed with mean $1$ and standard deviation
$\sqrt{\alpha}$. However, for the periodic bit sequence, Eqs.
(\ref{eins}) and (\ref{zwei}), the values of $\sigma^{\nu}$ and
$\xi^{\nu}_j$ are taken from the random bits $S_i$. For instance
$\sigma^{\nu}$ is identical with $\xi^{\mu}_j$ for $j=1, .., N$
and $\mu = \nu + N + 1 - j$. Taking this into account we find
for $1<\alpha<2$:
\begin{equation}
\label{acht} 
\begin{array}{ccl}
\langle E \rangle &  = & 
\cases {1+\frac1N&for $P$ even\cr1&for $P$ odd\cr}
\\[.5cm]
\langle E^2 \rangle &  =  & 
\cases{2+\alpha+\frac{6-\alpha}N-\frac4{N^2}&for $P$ even\cr
2+\alpha-\frac2N&for $P$ odd\cr}
\end{array}
\end{equation}

For $\alpha > 2$ the results above for odd $P$ hold for even
ones, too. Hence for $N \rightarrow \infty$ the standard
derivation of the $E^{\nu}$ values is $\sqrt{1+\alpha}$ instead of
$\sqrt{\alpha}$ of the uncorrelated case. The correlations
increase the noise relatively to the signal. Assuming a
Gaussian distribution of the $E^{\nu}$ values in the limit $N
\rightarrow \infty$, which is supported by our numerical
simulation, we obtain the training error $\varepsilon_t$ as 
\begin{equation}
\label{neun} \varepsilon_t = \phi \left( -
\frac{1}{\sqrt{1+\alpha}} \right) 
\end{equation}
with the error function
\begin{equation}
\label{zehn} \phi (x) = \int\limits^{x}_{-\infty} \,
\frac{1}{\sqrt{2\pi}} \, e^{-\frac{y^{2}}{2}} dy \,.
\end{equation}

If the random bits are not repeated periodically, but arranged
linearly as discussed above, the moments depend on the number
$\nu$ of the pattern. If $\nu = 1$ is the first and $\nu = P$ is
the last pattern, we define
\begin{equation}
\label{elf} \gamma = \left\{
\begin{array}{ccc}
\nu/N & \mbox{ for } & \nu \le N \\
1 & \mbox{ for } & \nu > N 
\end{array}
\right.
\end{equation}

In this case the training error depends on $\gamma$ and we find
\begin{equation}
\label{zwoelf} \varepsilon_t = \phi \left( -
\frac{1}{\sqrt{\alpha + \gamma^2}} \right) \,.
\end{equation}

Figure 3 shows the training error $\varepsilon_t (\alpha)$ for the
uncorrelated bits and the periodic bit sequence. In
the latter case $\varepsilon_t$ is averaged over the patterns.
The correlations of the bit sequence increase the training
error, in agreement with the decrease of the storage capacity
shown in the previous section.

\section {General Boolean function}

Up to now we have restricted our map to a perceptron. We
expect that multilayer networks can reproduce a larger bit
sequence, in accordance to the higher storage capacity of the
committee machine (Priel \etal 1994). In this section we study the
storage capacity of a general Boolean function $b : \{+1,-1\}^N
\rightarrow \{+1,-1\}$, which is the size of the random bit
sequence with period $P$ which can be reproduced by any Boolean
function $b$, i.e. 
\begin{equation}
\label{dreizehn} b(S_{\nu}, \dots, S_{\nu + N - 1}) = S_{\nu + N}
\, ; \, \nu = 1,\dots,P \,.
\end{equation}

Since we have the freedom to choose for any input configuration
$(S_{\nu},\dots, S_{\nu + N - 1})$ an arbitrary output bit $S_{\nu
+ N}$, our problem reduces to the question if all of the input
configurations are different from each other. If all $(S_{\nu},
\dots, S_{\nu + N - 1})$ are different then we can define a
Boolean
function which maps each of those states to the corresponding
bit $S_{\nu +N}$. For the rest of the $2^N - P$ input states we
have the freedom to choose an arbitrary output bit; hence in
this case, there
are $2^{(2^{N}-P)}$ many Boolean functions which map the bit
sequence correctly.

If two of the input configurations $(S_{\nu}, \dots, S_{\nu + N-
1})$ are identical there is still a probability of $1/2$ that
the
two output bits are different, too. To get an analytic estimate
for the size of a random bit sequence which can be reproduced by
a Boolean function we neglect correlations between the input
configurations. That means we consider $P$ configurations
$(S_1^{\nu},\dots, S_N^{\nu}) \, ; \, \nu = 1, \dots, P$ where all of
the bits $S^{\nu}_i$ are chosen randomly and independently. We
want to calculate the probability $f$ that all of the $P$ states
are pairwisely different. There are $2^N$ many possible states.
The first configuration $\nu = 1$ can be any of those states.
The second one can take any of the $2^N -1$ remaining states,
etc. Hence, the number $C$ of allowed configurations is
\begin{equation}
\label{vierzehn} C= 2^N(2^N - 1)(2^N -2) \cdots (2^N - P+1)
\end{equation}
which gives
\begin{eqnarray}
\label{fuenfzehn} \ln C  &=  \sum\limits^{P}_{\nu = 1} \, \ln
(2^N -\nu+1) = \sum\limits^{P}_{\nu =1} \left[ N\ln 2 + \ln \left(
1 - \frac{\nu-1}{2^{N}} \right) \right] \\
&= PN\ln 2 + \sum\limits^{P}_{\nu =1} \, \ln \left( 1 - 
\frac{\nu-1}{2^{N}} \right) \,.
\end{eqnarray}

If $P \ll 2^N$ we can expand $\ln$ and obtain
\begin{equation}
\label{sechzehn} \ln C \simeq PN\ln2 - \frac{1}{2^{N}} \,
\frac{P(P-1)}{2} \,.
\end{equation}

Since the total number of all possible configurations is
$2^{PN}$, the function of the allowed ones is
\begin{equation}
\label{siebzehn} f\simeq \exp \left[ - \frac{P(P-1)}{2^{N+1}}
\right].
\end{equation}

We define the average period $P_c$ by $f(P_c) =1/2$ and obtain
for large $N$
\begin{equation}
\label{achtzehn} P_c = \sqrt{2\ln2} \,\,\, 2^{\frac{N}{2}} \,.
\end{equation}

Hence we expect that the average length of the bit sequence
which can be reproduced by a Boolean function scales as the
square root of $2^N$. In fact our problem is similar to the
random map, where the average cycle length has the same scaling
property (Harris 1960, Derrida \etal 1987).

The configurations taken from a random bit sequence are
correlated, since consecutive configurations are obtained by
shifting a window of $N$ bits over the sequence. However, our 
numerical simulations show that these correlations do not 
change the scaling law Eq.(\ref{achtzehn}). 
For a given sequence with $P$ bits the size $N$ of
the window is increased until this sequence can be reproduced by
a Boolean function. $N_c$ is defined as the window size $N$
where 50\% of the sequences are reproduced. In Figure 4, $P$ is
shown as a function of $N_c$. For $P \le 17$, $N_c$ is determined
by exhaustive enumeration. For larger $P$ values $N_c$ is
estimated from up to $10^5$ random samples. The log-linear plot 
shows that the data are consistent with
\begin{equation}
\label{neunzehn} P = 1.6 \times \sqrt2^{N_c}\,.
\end{equation}

The comparison with Eq.(\ref{achtzehn}) shows that the
correlations seem to change the prefactor from $1.17$ to $1.6$,
but the number still increases with the square root of $2^N$, the
size of the input space.

\section{Summary}
A perceptron of $N$ input bits has been trained by a random
bit sequence with a period $P$. Each output bit is contained
in $N$ input vectors. These correlations decrease the storage capacity
to $\alpha_c=1.7\pm0.02$ compared to $\alpha_c=2$ for uncorrelated output 
bits. For the corresponding bit generator the bit sequence has a tiny 
basin 
of attraction.

An analysis of Hebbian weights shows that a bit sequence gives a larger 
noise to signal ratio than a random classification problem.
This result is in agreement with the lower storage capacity.

If a general Boolean function is trained by the random bit sequence, the
maximal period $P$ scales as the square root of $2^N$, the size of the 
input space.

\section*{Acknowledgements}
This work has been supported by the Deutsche Forschungsgemeinschaft
and the MINERVA center of physics of the Bar-Ilan University.
We thank Georg Reents for valuable discussions.
\vfill\eject
\References
\item[] Bork A 1994 {\it Zeitreihenanalyse} Diploma thesis 
(W\"urzburg: Institut f\"ur Theoretische Physik der Universit\"at 
W\"urzburg)
\item[] Cover T M 1995 {\it IEEE Trans. Electron. Comput.} {\bf EC-14} 326
\item[] Derrida B and Flyvbjerg 1987 \JP {\bf 48} 971 
\item[] Eisenstein E, Kanter I, Kessler D A and Kinzel W  1995 \PRL {\bf 
74} 6 
\item[] Fontanari JF and Meir R 1989 \JPA {\bf 22} L803
\item[] Gardner E 1988 \JPA {\bf 21} 257
\item[] Harris B 1960 {\it Ann. Math. Stat.} {\bf 31} 1045
\item[] Hertz J, Krogh A and Palmer R G 1991 {\it Introduction to the 
theory of neural computation} (Redwood City, CA: Addison Wesley) 
\item[] Kanter I, Kessler D A, Priel A and Eisenstein E 1995 \PRL {\bf 75} 
2614
\item[] Kinzel W and Opper M 1991 in {\it Models of Neural Networks} eds. 
Domany E, van Hemmen J L and Schulten K (Berlin: Springer) p 149
\item[] L\'opez B, Schr\"oder M and Opper M 1995 \JPA {\bf 28} L447
\item[] Monasson R 1992 \JPA {\bf 25} 3701
\item[] Opper M 1988 \PR A {\bf 38} 3824
\item[] Opper M and Kinzel W 1996 in {\it Models of Neural Networks III} 
eds. 
Domany E, van Hemmen J L and Schulten K (New York: Springer) p 151
\item[] Priel A, Blatt M, Grossmann T and Domany E 1994 \PR E {\bf 50} 577
\item[] Schl\"afli L 1950 in {\it Ludwig Schl\"afli 1814--1895: Gesammelte
Mathematische Abhandlungen} ed. Steiner-Schl\"afli-Komitee 
(Basel: Birkh\"auser) 171
\item[] Tarkowski W and Lewenstein M 1993 \JPA {\bf26} 2453
\item[] Weigand A S and Gershenfeld N A (eds) 1993 {\it Time Series 
Prediction, Forecasting the Future and Understanding the Past} 
(Santa Fe: Santa Fe Institute) 
\endrefs

\vfill\eject

%
%



\section*{Figure 1}
\def\spa{\hskip .41cm}
\def\spb{\hskip .0cm}
\begin{figure}
\def\spa{}
\leftline{\hskip4.9cm$S_\nu\hskip .05cm S_{\nu+1}\hskip 1.3cm
\hbox to0.8cm{\hss$\dots$\hss}S_{\nu+N-1}\spa S_{\nu+N}$}
\vskip0.2cm
\centerline{
\vbox to 2.5cm{
\hbox to 0cm{\strut\hss}\vss
\hbox to 0cm{\hskip5cm$\vec w=(w_1,\dots,w_N)$\hss}\vss
\hbox to 0cm{\hskip5cm$\sigma_\nu=\sign(\sum_{j=1}^Nw_jS_{j-1+\nu})$\hss}}
\epsfxsize=9cm
\epsffile{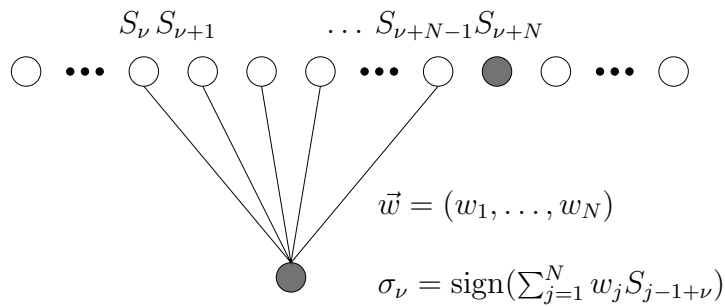}
}
\caption{A perceptron learning a periodic time
series. The desired output of the perceptron
(marked) is the next bit of the series and therefore part of other
input patterns as well.}
\end{figure}

\section*{Figure 2}
\begin{figure}
\centerline{
\epsfysize 5cm
\epsfbox{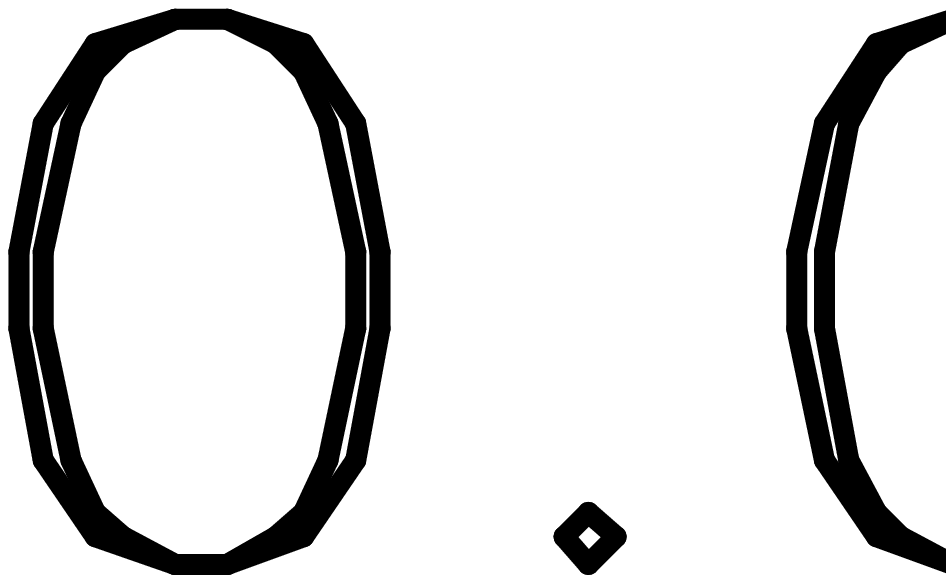}
\qquad
\epsfysize 5cm
\epsfbox{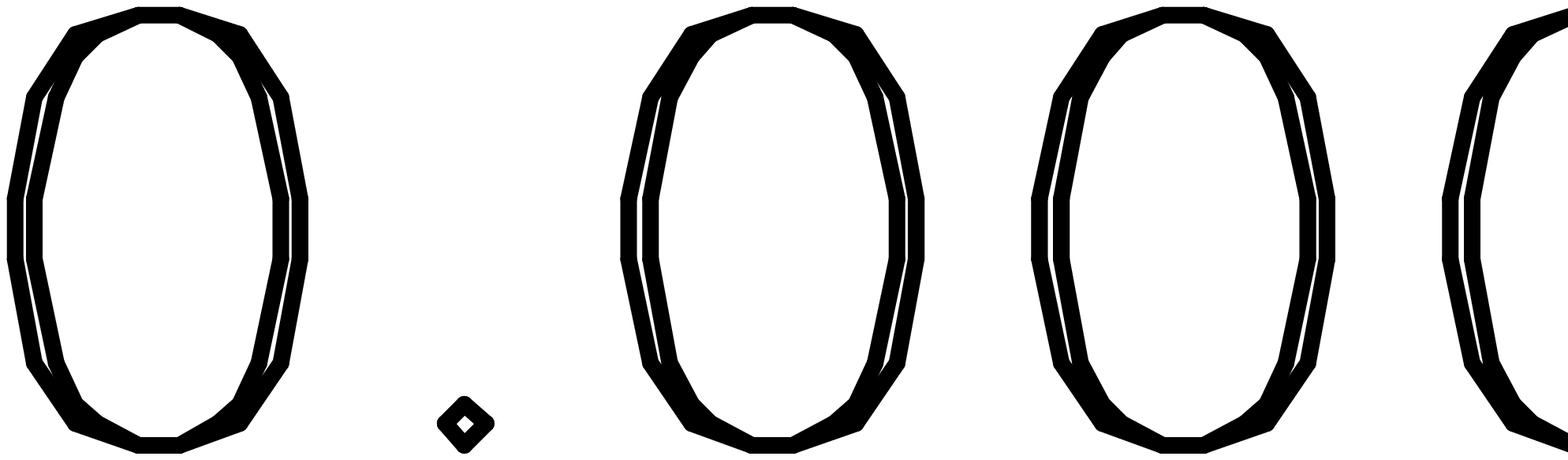}
}
\caption{
Left hand side: 
The probability $f$ of a bit sequence to be linearly separable as a 
function of $\alpha=P/N$. The sequence is constructed from $P$ random
bits which are repeated periodically. The simulations are performed for
a perceptron with $N=100$ input bits and $f$ is averaged over 50 sets of 
patterns at least.
Right hand side:
The median learning time to the power of $-1/2$ as
a function of $\alpha$. The size of the perceptron is $N=400$, and
$\tau$ is averaged over 1000 sets of patterns. The line is a least square fit
to the data.}
\end{figure}

\pagebreak

\section*{Figure 3}
\begin{figure}
\centerline{
\epsfysize 8cm
\epsfbox{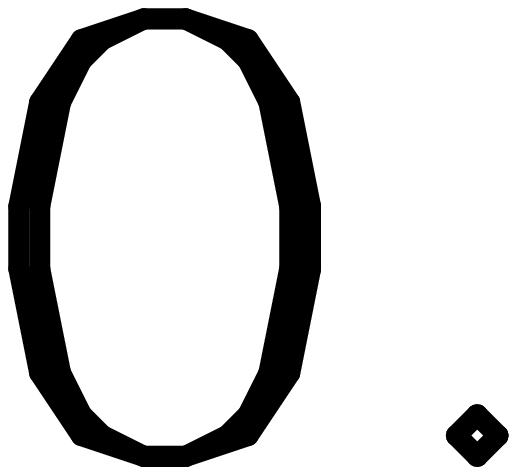}}
\caption{The training error of Hebbian weigths 
for different topologies. The inputs
are chosen binary. $\Diamond$: random patterns
and $\times$: patterns from a random bit sequence with periodic boundary
condition. The simulations were done for $N=200$ and averaged
over 100 samples each. The lines show the theoretical results.}
\end{figure}
\vfill\eject

\section*{Figure 4}
\begin{figure}
\vglue -4cm
\centerline{
\epsfysize 15cm
\epsfbox{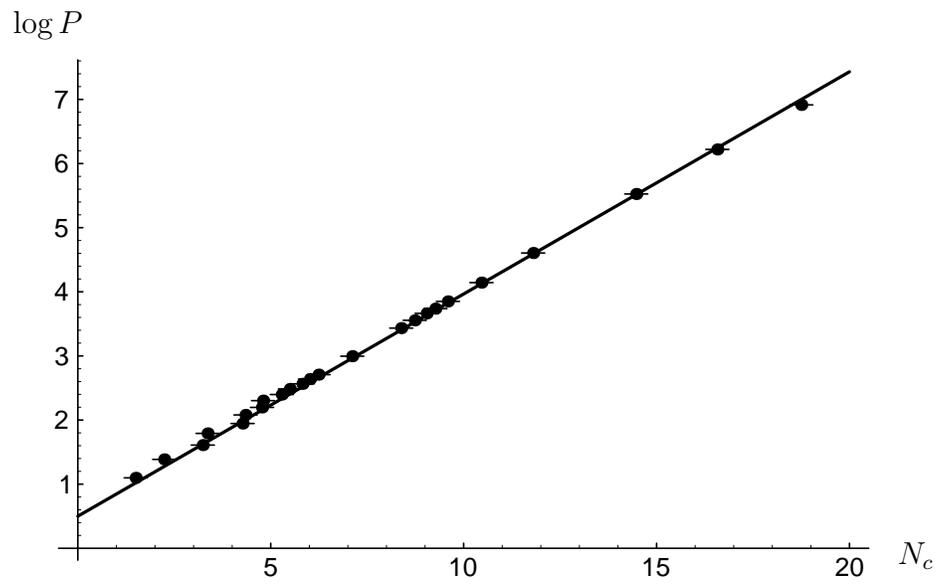}}
\vglue -4cm
\caption{The Length of a cycle that is learnable by a Boolean
function as a function of $N_c$.
The values up to $P=17$ are exact. The values up to $P=100$ are
averaged over 100000 samples, for $P=251$ over 50000, for $P=503$ over 1000 and
for $P=1007$ over 100 samples. The errorbars are given.
The line shows $P=\exp(0.5)\,\,2^{0.5N_c}$}
\end{figure}
\vglue -10.8cm \quad\quad\quad $\log P$
\vglue 6.5cm \hglue 13cm $N_c$

\vfill\eject
%
%

\begin{table}
\caption{The storage capacity of a perceptron learning different tasks.
Measured with (1) half-error and (2) median learning-time method }
\begin{indented}
\item[]\begin{tabular}{@{}lll}
\br
&method 1&method 2\\
\mr
random $N=100$&$1.99\pm0.01$&$1.995\pm0.01$\\
\mr
time series $N=100$&$1.85\pm0.025$&$1.82\pm0.02$\\
time series $N=400$&&$1.82\pm0.02$\\
\mr
ring $N=100$&$1.7\pm0.025$&$1.69\pm0.01$\\
ring $N=400$&&$1.7\pm0.02$\\
\mr
ring (rnd out) $N=100$&$1.98\pm0.05$&$1.99\pm0.01$\\
ring (rnd out) $N=400$&&$1.98\pm0.02$\\
\br
magnetization $m=0.4$&$N=100$&\\
\mr
ring&$1.95\pm0.05$&$1.95\pm0.03$\\
random&$2.25\pm0.05$&\\
\br
\end{tabular}
\end{indented}
\end{table}

\end{document}